\documentclass[twocolumn,showpacs,aps,prl,superscriptaddress]{revtex4}

\usepackage{graphicx}
\usepackage{dcolumn}
\usepackage{amsmath}
\usepackage{epsfig}

\input pubboard/babarsym

\def\Dstarp     {\ensuremath{D^{*+}}\xspace}
\def\Dstarm     {\ensuremath{D^{*-}}\xspace}

\def\Dp         {\ensuremath{D^+}\xspace}
\def\Dm         {\ensuremath{D^-}\xspace}

\def\DeltaEStd  {\ensuremath{\Delta E} \xspace}

\def\masslik    {\ensuremath{{\cal L}_{\rm Mass}}\xspace}

\newcommand{\BABARPubYear}    {03}
\newcommand{\BABARPubNumber}  {015}

\newcommand{\SLACPubNumber} {9986}

\def\figurebox#1#2#3{%
    \def\arg{#3}%
    \ifx\arg\empty
    {\hfill\vbox{\hsize#2\hrule\hbox to #2{\vrule\hfill\vbox to #1{\hsize#2\vfill}\vrule}\hrule}\hfill}%
    \else
    {\hfill\epsfbox{#3}\hfill}%
    \fi}

\begin{document}

\preprint{\babar-PUB-\BABARPubYear/\BABARPubNumber} 
\preprint{SLAC-PUB-\SLACPubNumber} 
\begin{flushleft}
\babar-PUB-\BABARPubYear/\BABARPubNumber\\
SLAC-PUB-\SLACPubNumber\\
\end{flushleft}

\title{
{\large  \boldmath
Measurement of Time-Dependent \CP Asymmetries and the \CP-Odd Fraction
in the Decay \Bztodstdst}
}
%
\author{B.~Aubert}
\author{R.~Barate}
\author{D.~Boutigny}
\author{J.-M.~Gaillard}
\author{A.~Hicheur}
\author{Y.~Karyotakis}
\author{J.~P.~Lees}
\author{P.~Robbe}
\author{V.~Tisserand}
\author{A.~Zghiche}
\affiliation{Laboratoire de Physique des Particules, F-74941 Annecy-le-Vieux, France }
% ...
\collaboration{The \babar\ Collaboration}
\noaffiliation

\date{\today}

\begin{abstract}
We present a measurement of time-dependent $C\!P$
asymmetries and an updated determination of the $C\!P$-odd fraction in the decay
$B^0 \rightarrow D^{*+}D^{*-}$ using
a data sample 
of $88 \times 10^{6} B\bar{B}$ pairs collected 
by the {\mbox{\slshape B\kern-0.1em{\smaller A}\kern-0.1em
    B\kern-0.1em{\smaller A\kern-0.2em R}}} detector at the
PEP-II $B$ Factory at SLAC.
We determine the $C\!P$-odd fraction to be $0.063 \pm 0.055\stat \pm 0.009\syst$.
The time-dependent $C\!P$ asymmetry parameters ${\rm Im}(\lambda_+)$
and $|\lambda_+|$ are determined
to be $0.05 \pm 0.29\stat \pm 0.10\syst$ and $0.75 \pm 0.19\stat \pm 0.02\syst$, 
respectively.  The 
Standard Model predicts these parameters to be $-\sin\! 2 \beta$ and 1, respectively, in
the absence of penguin diagram contributions.

\end{abstract}


\maketitle

\label{sec:Introduction}
The symmetry for combined charge conjugation {\it (C)} and parity {\it (P)} transformations 
is violated in $B$ decays. Measurements of \CP asymmetries by the
\babar~\cite{babarCP} and BELLE~\cite{belleCP} collaborations
established this effect and are compatible with
the Standard Model expectation based on the current knowledge of 
the Cabibbo-Kobayashi-Maskawa~\cite{CKM} quark-mixing matrix.
As a result of the interference between direct $B$ decay 
and decay after flavor change, a \CP -violating  asymmetry is expected in the
time evolution of the decays
$\Bz \to D^{*+} D^{*-}$~\cite{conjugates} 
within the framework of the Standard Model~\cite{angular}.
This \CP asymmetry is related 
to \stwob when corrections due to theoretically uncertain penguin diagram
contributions are neglected~\cite{gronau,sanda}.
Penguin-induced corrections are predicted to be small
in models based on the factorization approximation 
and heavy-quark symmetry; an effect of about $2\%$ is predicted by Ref.~\cite{xing}. 
A comparison of measurements of \stwob from 
$b \to c \bar{c} s$ modes such as 
$B^0 \to J/\psi K^0_S$~\cite{babar-stwob-newprl} with that 
obtained in \Bztodstdst is an important test of these models and
the Standard Model.

The \Bztodstdst mode is a pseudoscalar decay to a vector-vector final state,
with contributions from three partial waves with different \CP
parities: even for the $S$- and $D$-waves, odd for the $P$-wave.
The \CP-odd contribution
is predicted to be about $5.5\%$ in Ref.~\cite{xing2}. 
We present an updated~\cite{ref:dstdstprl} determination
of the \CP-odd fraction, $R_\perp$,
based on a one-dimensional time-integrated angular analysis.  
We also present a measurement of 
the time-dependent \CP asymmetry, obtained from 
a combined analysis of the time dependence 
of flavor-tagged decays
and the one-dimensional angular distribution of the decay products.
The data used in this analysis were collected with the \babar\ detector
at the \pep2 storage ring.  
The \babar\ detector
is described in detail elsewhere~\cite{ref:babar}. 
The data sample corresponds to about
$88 \times 10^6 $
$e^+e^- \to \Upsilon(4S) \to B\bar{B}$ events.

\label{sec:EventSelection}
\Bz mesons are exclusively reconstructed by combining two charged $D^{*}$
candidates reconstructed in the modes
$\Dstarp\to\Dz\pip$ and $\Dstarp\to\Dp\piz$. 
We include the $\Dstarp\Dstarm$ combinations 
$(\Dz\pip, \Dzb\pim)$ and $(\Dz\pip, \Dm\piz)$, 
but not $(\Dp\piz,\Dm\piz)$ due to 
the smaller branching fraction and larger backgrounds.
Prior to forming a \Bz,
the \Dstar candidates are subjected to a mass-constrained fit 
and vertex fit 
that includes the position of the beam spot.

The reconstructed \Dz and \Dp modes are $\Dz \to \Km \pip,\; \Km \pip
\piz, \; \Km \pip \pip \pim, \; \KS \pip \pim$, and $\Dp \to
\Km \pip \pip, \; \KS \pip, \; \Km \Kp \pip$.
The reconstructed mass of
the \Dz (\Dp) candidates is required to be
within 20\,\mevcc of the nominal \Dz (\Dp) mass~\cite{pdg},
except for $\Dz \to \Km\pip\piz$,
which has a looser requirement of 35\,\mevcc .
The $D$ candidates are subjected to a mass-constrained 
fit prior to forming \Dstar candidates.

Charged kaon candidates are required to be inconsistent with the pion
hypothesis, as inferred from the Cherenkov angle
measured by the Cherenkov detector
and the specific ionization measured by the charged-particle tracking system.  
No particle identification requirements are made
for the kaon from the decay $\Dz \to \Km \pip$.
The reconstructed mass of $\KS \to \pip\pim$ candidates is required 
to be within 25\,\mevcc of the nominal \KS mass.  The angle between the
flight direction and the momentum vector of the \KS is
required to be less than 200\,\mrad, and the transverse flight distance
from the primary event vertex must be greater than 2\,mm.
A mass-constrained fit is applied to each $\KS$ candidate.
Neutral pion candidates are formed from
two photons detected in the electromagnetic calorimeter,
 each with energy above 30\,\mev;
the mass of the pair must be within 20\,\mevcc\ of the 
nominal $\pi^0$ mass,
and their summed energy must be greater than 200\,\mev.
A mass-constrained fit is applied to these $\pi^0$ candidates. 
The mass of the $\pi^0$ from $D^{*+} \to D^+ \piz$,
however, is required to be within
35\,\mevcc\ of the nominal $\pi^0$ mass, and the momentum in the \FourS frame
in the interval $70 < |p^*| < 450\,\mevc$, with no requirement on the 
photon energy sum.

We construct a mass likelihood \masslik that includes 
the mass and mass uncertainty of the $D$ and \Dstar candidates.
The $D$ mass resolution is modeled by a Gaussian whose variance 
is determined on a candidate-by-candidate basis.
The $\Dstar$ -- $D$ mass difference resolution is modeled 
by a double-Gaussian distribution
whose parameters are determined from simulated events.
The value of $\masslik$ is used to select \Bz candidates,
with a different requirement used for each $D$ decay mode
combination.
In an event where more than one \Bz candidate is reconstructed,
the candidate with the largest $\masslik$ value is chosen.

The primary variables used to distinguish signal from background are
the energy-substituted mass,
$\mes \equiv \sqrt{E_{\rm Beam}^2 - p_B^2}$,
and the difference of the $B$ candidate energy from the beam energy,
$\DeltaEStd \equiv E_{B} - E_{\rm Beam}$,
where all variables are evaluated in the \FourS center-of-mass frame. 
The \Bz candidates are required to have $-39 < \DeltaEStd < 31\,\mev$ and
$\mes > 5.2\,\gevcc$.

To reject backgrounds from the $\epem \to \ccbar$ continuum 
process, events are required to have a ratio of second to zeroth Fox-Wolfram
moments~\cite{ref:fox} of less than 0.6.
We also require
that the cosine of the angle between the thrust axis of the reconstructed $B$
and the thrust axis of the rest of the event be less than 0.9.

After all selection criteria have been applied,
a fit to the \mes distribution using a Gaussian and an ARGUS
function~\cite{argus} for the signal and background, respectively,
results in a signal yield of $156 \pm 14 \stat$ 
events.  In the region $\mes > 5.27\, \gevcc$,
the signal purity is 73\%.

\label{sec:TransversityAnalysis}
We perform a one-dimensional angular
analysis to determine 
the fraction, $R_\perp$, of the $P$-wave, \CP -odd component
of the \Bztodstdst decay.  
In the transversity basis~\cite{angular}, the following three
angles are defined:
the angle $\theta_1$ between the momentum of the  
slow pion from the $D^{*-}$ in the $D^{*-}$ rest frame
and the direction of flight of the $D^{*-}$ in the $B$ rest frame;
the polar angle $\theta_{\rm tr}$ between the normal
to the $D^{*-}$ decay plane and the 
direction of flight of the slow pion from the $D^{*+}$ in the $D^{*+}$ rest frame; and 
the corresponding azimuthal angle $\phi_{\rm tr}$. 
The time-dependent angular distribution of the decay products 
is given in Ref.~\cite{penguin}.

The dependence of the detector efficiency on the decay angles can introduce a bias
in the measured value of $R_\perp$.
Including the 
efficiency explicitly in the decay rate and then integrating 
over time and the angles $\theta_1$ and $\phi_{\rm tr}$ results in the 
one-dimensional differential
decay rate:
  \begin{eqnarray}
\frac{1}{\Gamma}  \frac{d \Gamma}{d \cos \theta_{\rm tr}} & = & 
  \frac{9}{32\pi} \left[ (1-R_\perp) \sin^2 \theta_{\rm tr}  \right. \nonumber \\
& \times & \left\{ \frac{1+\alpha}{2} I_{0}(\cos\theta_{\rm tr}) +  
\frac{1-\alpha}{2} I_{\parallel}(\cos\theta_{\rm tr}) 
\right\} \nonumber \\
& + & \left. 2 R_\perp \cos^2 \theta_{\rm tr}  
\times I_{\perp}(\cos\theta_{\rm tr}) \right],
\label{AngDisArt}
 \end{eqnarray}
where 
$R_\perp = M_\perp^2 / (M_0^2 + M_\parallel^2 + M_\perp^2)$,
$\alpha = (M_{0}^2  - M_{\parallel}^2) / (M_0^2 + M_{\parallel}^2)$,
and $(M_0, M_\parallel, M_\perp)$ are the magnitudes of the amplitudes
in the transversity basis.
The three efficiency moments, $I_k$ $(k=0, \parallel, \perp)$, are defined as
\begin{eqnarray}
\displaystyle
I_{k}(\cos\theta_{\rm tr}) = \int \! d \! \cos \theta_1 \, d \phi_{\rm tr}
\; g_k(\theta_1, \phi_{\rm tr}) \,\epsilon(\theta_1,\theta_{\rm tr},\phi_{\rm tr}),
\label{moments}
\end{eqnarray}
where $g_0 = 4\cos^2\theta_1 \cos^2\phi_{\rm tr}$, $g_{||} = 2\sin^2\theta_1 \sin^2\phi_{\rm tr}$,
$g_{\perp} = \sin^2\theta_1 $, and $\epsilon$ is the detector efficiency.
The efficiency moments are determined using 
simulated events.
The efficiency moments are 
fit to second-order even polynomials in $\cos\theta_{\rm tr}$,
the parameters of which are fixed in 
the subsequent likelihood fit to the $\cos\theta_{\rm tr}$ distribution.

The measurement of $R_\perp$ is based on a combined 
unbinned maximum likelihood fit of the $\cos\theta_{ \rm tr}$ 
and \mes distributions.  The probability density function (pdf) for the 
\mes distribution is given 
by the sum of ARGUS and Gaussian functions.
The background shape
is modeled by an even second-order polynomial in $\cos\theta_{\rm tr}$.
The pdf used for signal events is given by Eq.~\ref{AngDisArt}. 
The experimental resolution of $\theta_{\rm tr}$
is not negligible and is accounted for by convolving the signal pdf
with a double Gaussian.
Also, the resolution of $\theta_{\rm tr}$ 
has significant tails
caused by mis-reconstructed events.  The effect of these tails
is accounted for
by an additional term in the signal pdf.  The parameterization of
the $\theta_{\rm tr}$ resolution is determined from simulations.

\begin{figure}[!ht]
\begin{center}
\epsfig{file=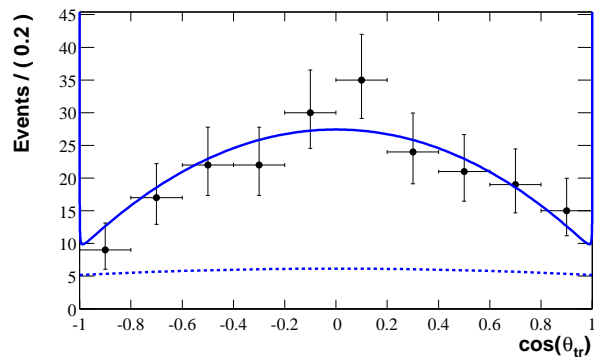,width=0.95\linewidth}
\caption{ Measured distribution of $\cos\theta_{\rm tr}$ and fit results. 
The data points are from the region $\mes > 5.27\,\gevcc$ and the solid line
is the projection of the fit result in the same region.  The dotted line
represents the background component.
}
\label{fig:datafit}
\end{center}
\end{figure}

We categorize our events in three types:
$\Dstarp\Dstarm \to (\Dz\pip,\Dzb\pim)$,  
$(\Dz\pip,\Dm\piz)$, and $(\Dp\piz,\Dzb\pim)$
because events with a
neutral slow pion and events with a charged slow pion have different
background levels, detection efficiencies, and 
$\cos\theta_{\rm tr}$ resolutions.
Thus, the parameters determined in the likelihood fit are three signal
fractions, the $\cos \theta_{\rm tr}$ background shape parameter, 
three \mes parameters ($\sigma$ and mean of the Gaussian, and $\kappa$ from 
the ARGUS function), and $R_\perp$.
The fit to the dataset yields a 
value of 
\begin{equation}
\displaystyle
R_\perp = 0.063 \pm 0.055\stat \pm 0.009\syst. 
\end{equation}
Figure~\ref{fig:datafit} shows the distribution of $\cos\theta_{\rm tr}$ for
events in the 
range $\mes > 5.27\,\gevcc$.
The value of $\alpha$ is fixed to 
zero in the fit,
incurring a (negligible) systematic uncertainty.
The largest systematic uncertainties
arise from
the parameterization of the angular resolution (0.005)
and the determination of the efficiency moments (0.005).

\label{sec:CPasymmetryAnalysis}

In addition to the time-independent measurement of the \CP-odd fraction,
we perform a combined analysis of the $\cos \theta_{\rm tr}$ distribution and 
the time dependence in order to determine
the time-dependent \CP asymmetry,
using the sample of \Bztodstdst events
described previously.
We also use information from the other $B$ meson in 
the event to tag its flavor as either a \Bz or \Bzb.

Although factorization models predict a small
penguin contamination in the weak phase difference 
in ${\rm Im}(\lambda_{f})=-\sin\!2\beta$~\cite{xing},
a sizable penguin contribution cannot {\it a priori} be excluded.
Thus, the value of 
$\lambda_{f}=\eta_{\CP} \frac{q}{p}\frac{\bar{A}(f)}{A(f)}$~\cite{penguin}
can be different for the three transversity amplitudes ($f = \perp,0,\parallel$)
because of possible different
penguin-to-tree ratios. 
This possibility is explicitly included in the 
parameterization of the decay rates described here.

The decay rate $F_+ (F_-)$ for a neutral $B$ meson tagged as a $B^{0}
(\Bzb)$ is given by
\begin{eqnarray}
F_\pm(\deltat) & = &
\frac{{\rm e}^{ - | \deltat |/\tau_{B^0} }}{4\tau_{B^0}}
 \Bigl\{ G(1-{\textstyle{\frac 12}}\Delta {\cal D}) \mp \nonumber \\
& &  {\cal D} \left[
 S \sin{ (\deltamd  \deltat) } 
+ C \cos{ (\deltamd  \deltat) }  \right]  \Bigr\},
 \label{eq:sincos}
\end{eqnarray}
where $\Delta t = t_{\rm rec} - t_{\rm tag}$ is the difference between 
the proper decay time of the reconstructed $B$ meson ($B_{\rm rec}$) and 
of the tagging $B$ meson ($B_{\rm tag}$),
$\tau_{\Bz}$ is the \Bz lifetime, and \deltamd 
is the mass difference determined from the \Bz-\Bzb oscillation frequency.
The dilution factor, ${\cal D} = 1 - 2\omega$,
where $\omega$ is the average mistag fraction,
describes the effect of incorrect tags,
and $\Delta {\cal D}$ accounts for possible differences in the mistag
probabilities for $\Bz$ and $\Bzb$.  
The $G$, $C$ and $S$ coefficients are defined as
\begin{eqnarray} 
& \!\!\!\!\!\!\!\!\!\!\!\!\!\!\!\!\!\!\!\!\!\!\!\!\!\!\!\!\!\!\!\!\!\!\!\!\!\!\!\!\!\!\!\! 
G = \, \frac{3}{4}  [(1-R_\perp) \sin^2\theta_{\rm tr} 
   +    2  R_\perp \cos^2\theta_{\rm tr}], \\
& \!\!\!\!\!\!C = \,\frac{3}{4} [(1-R_\perp)\frac{1 - |\lambda_{+}|^{2}}{1 +
  |\lambda_{+}|^{2}}  \sin^2\theta_{\rm tr}  +
 2 R_\perp \frac{1 - |\lambda_{\bot}|^{2}}{1 + |\lambda_{\bot}|^{2}}  \cos^2\theta_{\rm tr}],  \nonumber\\
&\!\!\!\!\!\!\!\!S \,= -\frac{3}{4}[(1-R_\perp)  \frac{2{\rm Im}(\lambda_{+})}
             {1 + |\lambda_{+}|^{2}} \sin^2\theta_{\rm tr} -
 2 R_\perp \frac{2{\rm Im}(\lambda_{\bot})}{1 +  |\lambda_{\perp}|^{2}} \cos^2\theta_{\rm tr} ]. \!\!\!\!\!\!
 \nonumber
\label{eq:ocs}
\end{eqnarray}
Because the two $\CP$-even transversity amplitudes produce the same
distribution in $\cos\theta_{\rm tr}$, we are only sensitive to $\lambda_+$,
the appropriate average of $\lambda_\parallel$ and $\lambda_0$:
\begin{eqnarray}
 \frac{{\rm Im}(\lambda_{+})}{1+|\lambda_{+}|^2} &=&
 \frac{
 \frac{{\rm Im}(\lambda_{\parallel})}{1+|\lambda_{\parallel}|^2}M_{\parallel}^2+\frac{{\rm Im}(\lambda_{0})}{1+|\lambda_{0}|^2}M_{0}^2}{M_{\parallel}^2 +M_{0}^2},
\nonumber \\ 
 \frac{1-|\lambda_{+}|^2}{1+|\lambda_{+}|^2}&=&
 \frac{
 \frac{1-|\lambda_{\parallel}|^2}{1+|\lambda_{\parallel}|^2}
 M_{\parallel}^2+ 
 \frac{1-|\lambda_{0}|^2}{1+|\lambda_{0}|^2} M_{0}^2 }
      {M_{\parallel}^2 +M_{0}^2}. 
\end{eqnarray}
If angular acceptance effects are not taken
into account and the \CP-odd fraction is allowed to float in the fit, 
then no bias is seen in the resulting value of $\lambda_+$ 
based on simulations.
Hence,
a dedicated method to correct for detector efficiency is not required.
The value of $R_\perp$ obtained is therefore an effective value,
which is not identical to the 
acceptance-corrected value from the time-integrated measurement.

The time interval \deltat is calculated
from the measured separation \deltaz between the decay vertex of the 
reconstructed  $B$ meson and the vertex of the
flavor-tagging $B$ meson along the collision axis. 
Events with a \deltat\ uncertainty $< 2.5\ps$, 
and a measured $\vert \deltat \vert < 20 \ps$ are accepted.  
The mistag fractions and $\Delta t$ resolution functions are
determined from a sample of neutral $B$ decays to
flavor eigenstates, $B_{\rm flav}$, 
as in the \stwob measurement using
charmonium decays~\cite{babar-stwob-newprl}.
Vertex reconstruction, the determination of $\Delta t$, and
the algorithms used for the determination of the flavor of $B_{\rm tag}$
are described in Refs.~\cite{babar-stwob-newprl,babar-sin2b-prd}.  

\begin{figure}[tp]
\begin{center}
\epsfig{figure=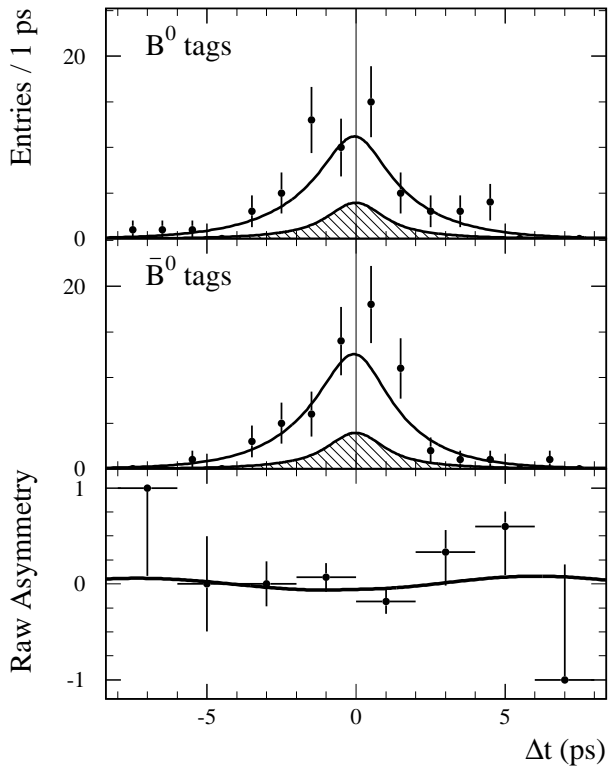,width=0.95\linewidth}
\caption{From top to bottom: Number $N_{\Bz }$ $(N_{\Bzb})$ of candidate 
events 
in the region $\mes > 5.27\,\gevcc$ with a \Bz (\Bzb) tag,
and the raw asymmetry
$(N_{\Bz}-N_{\Bzb})/(N_{\Bz}+N_{\Bzb})$, as functions of \deltat . The
solid curves represent the result of the combined fit 
to the full sample.
The shaded regions represent the background contributions.
}   
\label{fig:cpdeltat}
\end{center}
\end{figure}

We determine the parameters ${\rm Im}(\lambda_{+})$ and $|\lambda_{+}|$ with a simultaneous unbinned maximum likelihood fit 
to the \deltat distributions of the $B_{\rm rec}$ and $B_{\rm flav}$ tagged
samples (Fig.~\ref{fig:cpdeltat}). 
The \deltat distribution of the $B_{\rm flav}$ sample evolves
according to the known frequency for flavor oscillations in neutral $B$
mesons. The observed magnitude of the \CP asymmetry in the
$B_{\rm rec}$ sample and the flavor oscillation in the $B_{\rm flav}$ sample 
are reduced by the same factor ${\cal D}$ due to flavor mistags. The
\deltat distributions for the $B_{\rm rec}$ and $B_{\rm flav}$ samples are
both convolved with a common \deltat resolution function. 
The $\theta_{\rm tr}$ angular resolution
is accounted for in the same way as described previously.
Events are assigned signal and background probabilities based on  
their  \mes\ values.
Backgrounds are incorporated with an empirical
description of their \deltat distributions, containing prompt (zero
lifetime) and 
non-prompt components convolved with a separate resolution   
function~\cite{babar-stwob-newprl}.

A total of $38$ parameters are varied in the fit: the values of
${\rm Im}(\lambda_{+})$ and $|\lambda_{+}|$ (2), 
the effective \CP-odd fraction (1),
the average mistag fraction $w$ and the difference $\Delta w$
between $B^{0}$ and $\Bzb$ mistags for each tagging category (8),
parameters for the signal $\Delta t$ resolution (9),
and parameters for the background time dependence (7),
$\Delta t$ resolution (3), and mistag fractions (8).
Because the \CP-odd fraction is small, we have little sensitivity to
the parameters $|\lambda_{\perp}|$ and ${\rm Im}(\lambda_{\perp})$. 
Therefore they are fixed to
$1.0$ and $-0.741$~\cite{babar-stwob-newprl} respectively.  
These are the values expected
if direct \CP
violation and contributions from penguin diagrams are neglected. 
The changes in the fitted values of ${\rm Im}(\lambda_{+})$ 
and $|\lambda_{+}|$ for 
different input values of ${\rm Im}(\lambda_{\perp})$ (varied between
$-1.0$ and $1.0$) and $|\lambda_{\perp}|$
(varied between 0.7 and 1.3) are taken into account as systematic 
uncertainties.
The results obtained from the fit (Fig.~\ref{fig:cpdeltat}) are
\begin{eqnarray}
{\rm Im}(\lambda_+) & = & 0.05 \pm 0.29\stat \pm 0.10\syst\\
|\lambda_+| & = & 0.75 \pm 0.19\stat \pm 0.02\syst.
\end{eqnarray}

The dominant sources of systematic uncertainty come from 
the variation of
the value of $\lambda_{\perp}$ (0.056 and 0.008, respectively, for ${\rm
  Im}(\lambda_{+})$ and $|\lambda_{+}|$), and
the level, composition, and \CP asymmetry of the background (0.078 and
0.005).

If the $B \to D^{*+}D^{*-}$ transition proceeds only through the 
$b \to c \bar{c} d $
 tree amplitude, we expect that ${\rm Im}(\lambda_+) = -\sin2\beta$ and
 $|\lambda_+| =1.$
To test this hypothesis, we fix ${\rm
  Im}(\lambda_+)=-0.741$~\cite{babar-stwob-newprl}
 and $|\lambda_+|=1$ 
and repeat the fit. The observed change in the likelihood
corresponds to 2.5 standard deviations (statistical uncertainty only).

\label{sec:Summary}
In summary, we have reported a measurement of the \CP-odd fraction 
and measurements of time-dependent \CP
asymmetries for the decay
\Bztodstdst.  The measurement of $R_\perp$ 
supersedes the previous \babar\ result~\cite{ref:dstdstprl}, 
with a factor of three reduction in the statistical uncertainty, and 
indicates that \Bztodstdst is mostly \CP-even.
The time-dependent asymmetries
are found to differ slightly from Standard Model predictions 
with penguin amplitudes ignored.

We are grateful for the excellent luminosity and machine conditions
provided by our \pep2\ colleagues, 
and for the substantial dedicated effort from
the computing organizations that support \babar.
The collaborating institutions wish to thank 
SLAC for its support and kind hospitality. 
This work is supported by
DOE
and NSF (USA),
NSERC (Canada),
IHEP (China),
CEA and
CNRS-IN2P3
(France),
BMBF and DFG
(Germany),
INFN (Italy),
FOM (The Netherlands),
NFR (Norway),
MIST (Russia), and
PPARC (United Kingdom). 
Individuals have received support from the 
A.~P.~Sloan Foundation, 
Research Corporation,
and Alexander von Humboldt Foundation.

\end{document}